# Metamorphic Virus Variants Classification
# Using Opcode Frequency Histogram

BABAK BASHARI RAD, MASLIN MASROM
Advanced Informatics School, UTM Razak School of Engineering and Technology
University Technology of Malaysia International Campus
Jalan Semarak, 54100 Kuala Lumpur
MALAYSIA
bbabak3@live.utm.my, maslin@ic.utm.my

*Abstract*- In order to prevent detection and evade signature-based scanning methods, which are normally exploited by anti-virus softwares, metamorphic viruses use several various obfuscation approaches. They transform their code in new instances as look entirely or partly different and contain dissimilar sequences of string, but their behavior and function remain unchanged. This obfuscation process allows them to stay away from the string based signature detection. In this research, we use a statistical technique to compare the similarity between two files infected by two morphed versions of a given metamorphic virus. Our proposed solution based on static analysis and it uses the histogram of machine instructions frequency in various offspring of obfuscated viruses. We use Euclidean histogram distance metric to compare a pair of portable executable (PE) files. The aim of this study is to show that for some particular obfuscation methods, the presented solution can be exploited to detect morphed varieties of a file. Hence, it can be utilized by non-string based signature scanning to identify whether a file is a version of a metamorphic virus or not.

*Key-Words:* Computer Virus, Metamorphic Virus, Obfuscation Techniques, Virus Detection, Virus Classification, Opcode Frequency Histogram,

## 1 Introduction

Today, metamorphic viruses are one of the most challenging issues in the context of computer security. Most of current anti-virus detector engines are generally based on syntactic features of viruses. They usually scan sequences of binaries in machine code level of files to find the signature of viral code. Syntactic characteristics make the anti-virus scanner vulnerable against the smart virus writers who are increasingly employing mutation techniques that change the byte strings of viruses syntactically, while the function and behavior of the virus will remain significantly unchanged [1].

To overcome metamorphic viruses' obfuscation techniques, semantic based features become into new anti-virus researches in recent years. The most crucial concern related to semantic methods is that it requires a large amount of prerequisites works. It normally takes a great deal of time to form a proper semantic signature. In addition, the semantic methods cannot be used on-the-fly [2]. It is also considerable to notice that a method, which spends an unreasonable time to analyze and detect a mutated variation of a virus, is not practicable, even it be able to detect with high degree of accuracy.

On the other hand, syntax based method or string signature strategy, need an incessantly updated database of virus signatures. Therefore, to keep a reliable and powerful database, a large amount of time of experts should be spent to extract signatures and produce database with as the most accurate data as possible [3]. However, all these efforts are confronted by obfuscation techniques and may be defeated easily.

Metamorphic virus writers use different innovative morphing techniques to evade the signature-based analysis. Some common obfuscation skills used in this kind of viruses are 1) garbage code insertion, 2) register usage exchange, 3) instruction replacement, 4) instruction permutation and 5) Code Transposition [2,4,5,6,7].

In this paper, we use the histogram of instructions opcodes as a statistical feature to compare files. In this way, we attempt to find out that whether a file is a morphed version of another one or not. We believe that although obfuscation techniques apply many changes into shapes of different variants of a virus, there are still some essential common properties among these variants, which contain the key instructions indicating their correspondent functions. In other words, obfuscation techniques are not able to remove statistical likelihood properties of two codes, totally, that are functionally similar.

Next section presents some related works has been done, previously. In section 3, we review some of the most common obfuscation methods usually utilized by metamorphic virus authors. Our approach is introduced in section 4, and experiments and result are presented in section 5. Finally, section 6 gives the conclusion and a few recommendations for future works.

## 2  Related Works

In [8], Szor and Ferrie introduced a valuable definition of metamorphic viruses and evolution of the code. They also introduced some basic metamorphic virus detection methods, in general, following with many useful examples.

Konstantinou, in his technical report [5], gave a comprehensive and detailed explanation for metamorphic viruses, obfuscation techniques and other advanced skills normally used by them. Then, he discusses about metamorphic virus detection methods, briefly.

The method introduced in [1] is based on this concept that properties of malwares are positioned in their semantics. Preda et al in this paper recommended a semantics-based structure for malware detectors. Their approach uses trace semantics to distinguish the behaviors of malware while the program code is being inspected for infection.

A helpful explanation of computer virus strategies and detection methods is accessible in [9] by authors. They explained static and dynamic detection approaches, mechanism of metamorphic virus engine and open problems in computer anti-virus technologies.

In [2], Karnik et al presented a method based on frequency of instructions using cosine similarity analysis to detect obfuscated viruses.

Webster and Malcolm in [10] introduced an approach towards metamorphic computer virus detection by an algebraic specification of the IA-32 assembly programming language. Their proposed method based on a specification in OBJ of the IA-32 instructions.

## 3  Obfuscation Techniques

As mentioned in previous section, metamorphic viruses utilize different techniques to defeat string signature based detection. In fact, metamorphic virus is able reprogram itself to challenge deeper static analysis [9].

In following, we review some of popular obfuscation techniques with examples of morphed codes, to understand how obfuscation may change the sequence of bytes in an executable to neutralize scanning.

### 3.1 Garbage Code Insertion

The simplest technique used by metamorphic engine to change the byte sequence of viral code is garbage code (or dead code) insertion. Inserted instruction has no effect on function of the code. There are different kinds of garbage code insertion.

In Table 1, some sample dead codes are given [6], which are actually useless instructions and semantically equivalent to no operation (NOP).

Table 1: Examples of Dead Codes

| Instruction Rule | Operation |
|---|---|
| add Reg, 0 | Reg ← Reg + 0 |
| mov Reg, Reg | Reg ← Reg |
| or Reg, 0 | Reg ← Reg \| 0 |
| and Reg, -1 | Reg ← Reg & -1 |

None of the operation samples, in Table 1, changes value of the register. Adding a value 0 to a register or a variable, transferring a register value to itself, a logical OR operation of register or variable with a 0 and a logical AND operation of a register or variable with a same length register or variable filled by binary value 1, or immediate constant -1 will not affect on the execution process.

In another kind of garbage code insertion, programmer inserts an instruction, which may changes the situation of machine, but before it involves the execution; programmer undoes it by one or more other instructions. Two examples are listed in Table 2.

Table 2: Examples of Garbage Codes

| Garbage Instructions | Comments |
|---|---|
| push cx<br>pop cx | Before any effects, it returns the value to the register from stack |
| inc ax<br>sub ax, 1 | Value of ax remain unchanged |

However, more mixed and complicated techniques of garbage code insertion can be used in metamorphic viruses. The following example is a piece of the Win32.Evol virus [11]. Its metamorphic engine inserts junk instruction among the main instruction to change the byte string of the code. Two different offspring of Win32.Evol are shown in table 3 and Table 4:

Table 3: Version 1 of Win32.Evol

| Binary Opcode | Assembly Code |
|---|---|
| C7060F000055 | mov [esi], 5500000Fh |
| C746048BEC5151 | mov[esi+0004],5151EC8Bh |
| String Signature: | C7060F000055C746048BEC5151 |

Table 4: Version 2 of Win32.Evol (junk insertion)

| Binary Opcode | Assembly Code |
|---|---|
| BF0F000055 | mov edi,5500000Fh |
| 893E | mov [esi],edi |
| 5F | pop edi |
| 52 | push edx |
| B640 | mov dh,40 |
| BA8BEC5151 | mov edx,5151EC8Bh |
| 53 | push ebx |
| 8BDA | mov ebx,edx |
| 895E04 | mov [esi+0004],ebx |
| String Signature: | |
| BF0F000055893E5F52B640BA8BEC5151538DA895E04 | |

As it can be obviously seen in tables 3 & 4, these two versions of the virus are completely different in looking, but their functions are same. Both transfer two double words into memory address specified by `esi`. The most noticeable point is that it is not possible to find a common sequence of bytes in both to use as a signature string of virus, even utilizing wildcards.

## 3.2 Register/Variable Usage Exchange

Usage of different registers or memory variables is another simple transformation method that metamorphic engines use it to mutate their code. This technique attempt to evade the string signature based detection as well, by changing similar bytes in various generations. In December 1998, Win95.Regswap utilized it to create different variants of the virus. It is clear that it does not influence on the function of the code, but the sequence of binaries will alter. Two various versions of Win95.Regswap are shown in Table 5 and Table 6 [8,11].

Highlighted strings show the common bytes in two instances. The complexity of this method is not too high and such scanners that use wildcards are able to detect the variants of the virus easily, because there are enough similar byte sequences to extract a signature string.

Table 5: Version 1 of Win95.Regswap

| Binary Opcode | Assembly Code |
|---|---|
| 5A | pop edx |
| BF04000000 | mov edi,0004h |
| 8BF5 | mov esi,ebp |
| B80C000000 | mov eax,000Ch |
| 81C288000000 | add edx,0088h |
| 8B1A | mov ebx,[edx] |
| 899C8618110000 | mov [esi+eax*4+00001118],ebx |
| String Signature: | |
| 5ABF040000008BF5B80C00000081C2880000008B1A899C8618110000 | |

Table 6: Version 2 of Win95.Regswap

| Binary Opcode | Assembly Code |
|---|---|
| 58 | pop eax |
| BB04000000 | mov ebx,0004h |
| 8BD5 | mov edx,ebp |
| BF0C000000 | mov edi,000Ch |
| 81C088000000 | add eax,0088h |
| 8B30 | mov esi,[eax] |
| 89B4BA1811000 | mov [edx+edi*4+00001118],esi |
| String Signature: | |
| 58BB040000008BD5BF0C00000081C0880000008B3089B4BA18110000 | |

However, the combination of this technique with other methods such as dead code insertion can make new generations enough difficult to detect and make the syntax signature based detection entirely impossible.

## 3.3 Instruction Replacement

This obfuscation method actually substitutes some instructions with their equivalent instructions in newer copies. Sometimes, programmers can perform an action in different ways of coding. For example, to assign 0 to register `eax`, following codes are possible:

```
mov eax, 0
xor eax, eax
and eax, 0
sub eax, eax
```

Therefore, this is a great opportunity for virus programmers to utilize this possibility in metamorphic engines. This method is like using different synonyms in human language [2].

The following codes, in Tables 7 and 8, show two versions of W95.Bistro taken from [4].

Some instructions or actions replaced by their equivalents. "test esi, esi" replaced by "or esi, esi". Instruction "or edi, edi" replaced by "test edi, edi", and finally, "mov ebp, esp" replaced by two consequent instructions "`push esp`" and "`pop ebp`", which perform same action. As it is clear, these replacements transmute the sequence of instructions binary codes. Thus, the string signature for these two versions of W95.Bistro is dissimilar. Certainly, like the previous method, some parts of string signature are similar and wildcards can be used for detection. The identical bytes are colored in Tables 7 and 8.

Table 7: Version 1 of Win95.Bistro

| Binary Opcode | Assembly Code |
|---|---|
| 55 | push ebp |
| 8BEC | mov ebp, esp |
| 8B7608 | mov esi, dword ptr [ebp + 08] |
| 85F6 | test esi, esi |
| 743B | je 401045 |
| 8B7E0C | mov edi, dword ptr [ebp + 0c] |
| 09FF | or edi, edi |
| 7434 | je 401045 |
| 31D2 | xor edx, edx |
| String Signature: | |
| 55 8BEC 8B7608 85F6 743B 8B7E0C 09 FF 7434 31D2 | |

Table 8: Version 2 of Win95.Bistro

| Binary Opcode | Assembly Code |
|---|---|
| 55 | push ebp |
| 54 | push esp |
| 5D | pop ebp |
| 8B7608 | mov esi, dword ptr [ebp + 08] |
| 09F6 | or esi, esi |
| 743B | je 401045 |
| 8B7E0C | mov edi, dword ptr [ebp + 0c] |
| 85FF | test edi, edi |
| 7434 | je 401045 |
| 28D2 | sub edx, edx |
| String Signature: | |
| 5554 5D 8B7608 09 F6 743B 8B7E0C 85 FF 7434 28D2 | |

## 3.4 Instruction Permutation

In some pieces of code, it is possible to change the sequence of instructions with no disturbing the execution. Byte strings in different versions of the code will appear unlike via this disordering technique.

If there is no dependency among some instructions, they can be reordered. Consider the following instructions:

```
op1      Reg1, Reg2
op2      Reg3, Reg4
```

If the below conditions are satisfied, these two instructions can be substituted [7]:

```
1- Reg1 is not equal to Reg2
2- Reg1 is not equal to Reg4
3- Reg2 is not equal to Reg3
```

For example, codes in two columns of Table 9 are equivalent and can be swapped, simply.

Table 9: Example of Instruction Permutation

| Code Order 1 | Code Order 2 |
|---|---|
| mov  eax, 0F | add  esi, ebx |
| push ecx | mov  eax, 0F |
| add  esi, ebx | push ecx |

## 3.5 Code Transposition

This technique modifies the structure of the program in form of physically reordering of the program codes, while preserving the execution order or flow of the program running using conditional jumps or unconditional branches. It may be done at the level of instructions or modules.

Fig. 1 shows an example of such code structure modification used in Zperm virus [8].

# 4 Proposed Methodology

In some of the mentioned techniques, such as register/variable exchange or instruction permutation or even in some cases of code transposition, frequency of similar instructions is same in different generations of morphed viruses. Our proposed solution deals with the frequencies of opcodes used in variants as their features and measures the dissimilarity between two files according to these features. We expect that if the obfuscation engine utilizes some special kinds of morphing techniques, the frequencies of identical instructions are approximately similar.

In addition, to achieve a better comparison, we breakdown the files into their building subroutines and compare two files according to their function blocks.

We make an instruction frequency histogram for each code block or subroutine. Then we can evaluate dissimilarity of two blocks by measure the distance between their histograms. It can be done by different histogram distance measurements techniques introduced in data mining techniques. In fact, each subroutine of a program is presented by a histogram of the contained instructions as a feature, in form of a vector, which the length of vector is equal to the number of total instructions of the machine. If we were able to compare the histograms of the building blocks of two programs, then we will be able to calculate the dissimilarity between two programs using this feature.

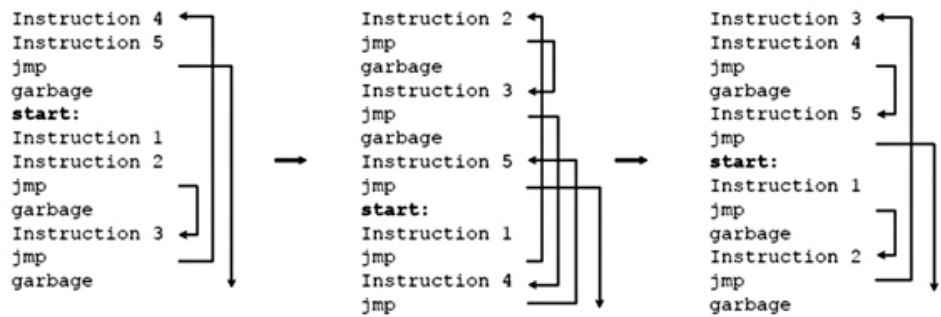

Fig. 1: Example of Code Transposition in different generations

Therefore, if dissimilarity value between two programs is less than a specific threshold we can conclude these two programs are morphed versions of each other. By this way, we can classify different variants of a metamorphic virus, which use some special types of obfuscation techniques. It is significant to mention that we can use this approach in the cases which the applied obfuscation has no effects or small changes on the frequencies of instructions.

## 4.1 Data Structure and Algorithm

To measure the dissimilarity between two executable files, we follow two general steps. In first step, is a pre-process, we prepare our input data in form of histograms as features. In the second step, which is a comparison process, we evaluate the dissimilarity of a pair of programs by comparison of their histograms.

In pre-process section, first, we disassemble executable files using IDA Pro 4.9 [12] and create assembly code files. Then, we analyze each assembly program and extract all procedures inside and save them as separate files. In next step, we create a set of histograms represent the frequencies of instructions within the procedures for each file. As a result, for each program, we will have a set of histograms, each one for a sub-procedure. Fig. 2 shows the process of program disassembly and breakdown into building subroutine blocks.

In the second step, we compare two programs by the use of their sets of histograms. Our comparison method is similar to that proposed in [2] with some changes to improve the algorithm. The comparison algorithm works as follow:

We have a set of histogram for each program. Each histogram driven for a subroutine inside the program and includes the frequency of each instruction in the subroutine.

Given two programs *P1* and *P2*, containing *m* and *k* subroutines, respectively. Therefore, *P1* has *m* has *m* histograms and *P2* has *k* histograms. Each histogram of *P1* is compared with all histograms of *P2*. According to our distance metric introduced in next section; distance value for each comparison will be calculated. More precisely, histogram $h_i$ from *P1* is compared with all histograms $h_j$, $1 \leq j \leq k$, from *P2*. That pair of histograms which has the minimum distance considered as the most similar histograms and consequently, we can consider their corresponding subroutines as mutated versions of each other. We save the minimum distance value for subroutine *i*. After all histograms of *P1* compared with all histograms of *P2*, we have a vector of length *m*, which contains the minimum distance values. We can use the average of this vector as the total distance value of *P1* and *P2*. It is important to take notice that this distance value is not symmetric. It means *distance(P1, P2)* is not equal to *distance(P2, P1)*. Hence, to get a more precise result, we define the distance of *P1* and *P2* as following:

$$d\{P1, P2\} = \frac{d(P1, P2) + d(P2, P1)}{2} \qquad (1)$$

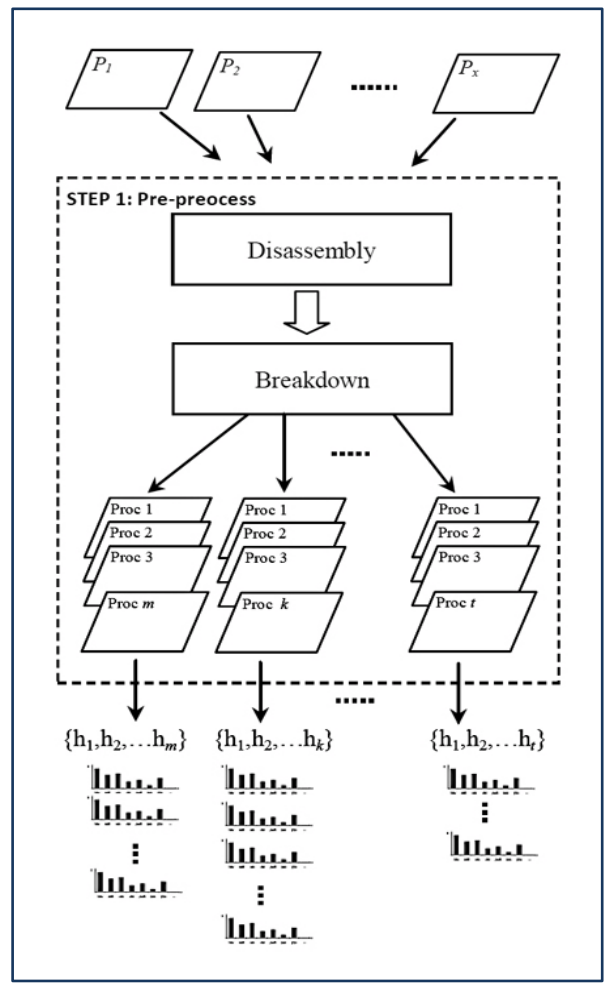

Fig. 2: Disassembly, breakdown and feature extraction

Fig. 3 shows the process of comparison between two programs *P1* and *P2*, briefly.

## 4.2 Dissimilarity Metric

There are various metric methods for measuring histogram dissimilarity and each of them has its application in related works. The first class of dissimilarity measurement is based on Minkowski-form distance metric [13]. Consider two vectors of size *n*, $X = (x_1, x_2, ..., x_n)$ and $Y = (y_1, y_2, ..., y_n)$, then the Minkowski-form distance between two vectors X and Y is calculated as:

$$d_{X,Y}^r = \sum_{i=1}^{n} |x_i - y_i|^r \qquad (2)$$

One of the most popular histogram distance measurements is Euclidean form distance. It is a Minkowski-form metric with *r* = 2, as following:

$$d_{X,Y} = \sum_{i=1}^{n} (x_i - y_i)^2 \qquad (3)$$

As shown in Fig. 4, histogram dissimilarity measures based on Minkowski-form compare only the parallel elements. It is appropriate for our case that the instructions opcodes are not related.

In addition, because we are going to test different kinds of programs in our data set, as we explained in 5.2, to obtain a common threshold for classification, we normalized the histograms before we begin to calculate the distance values as following:

$$X = \frac{X}{\sum_{i=1}^{m} x_i} \qquad (4)$$

Normalization of histogram helps us to find a program independent threshold for our data set.

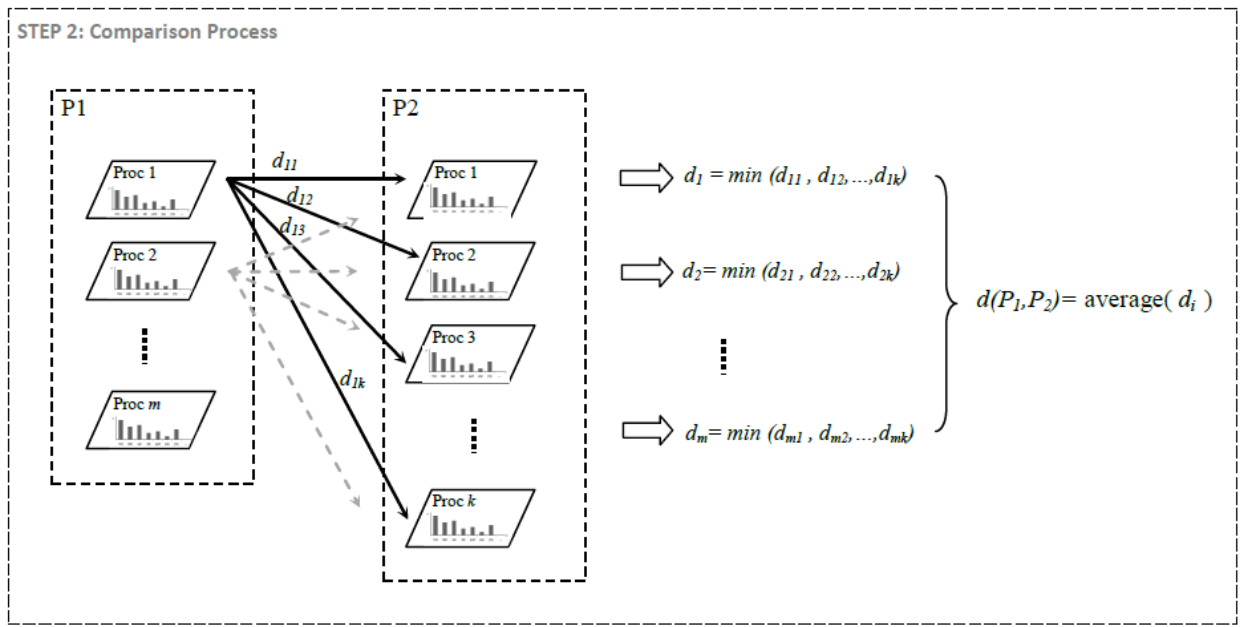

Fig. 3: Distance Calculation between two program *P1* and *P2*

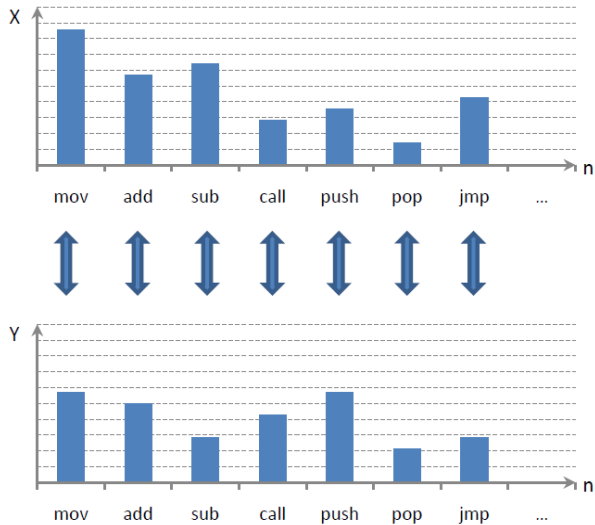

Fig. 4: Minkowski-form distance metrics compare only identical bins between two histograms.

## 5 Experiments and Findings

We used MathWorks MATLAB R2008a [14] to carry out our experiments include of preparing the data structure and implementing the comparison algorithm, and distance calculation.

### 5.1 Data Set

In our test data, we used several different obfuscated versions of some popular metamorphic viruses retrieved from [15] and a number of randomly chosen benign programs. Viruses and legal programs used in the experiment are listed in Table 10.

An in detail investigation and analysis of Win32.Evol is given in [16].

### 5.2 Results and Discussion

Table 11 shows the comparison results for each pair of two files. Because we use the average of distance for *P1* and *P2*, the table is symmetric.

The lower values indicate that programs are more similar. If the distance value is lower than a specified threshold, then we can conclude those two programs are obfuscated versions. Therefore, choosing the appropriate threshold value is critical. Choosing a low threshold value may cause to lose correct similar programs and over high threshold, value may produces false positive mistakes. We believe, in general, threshold value must be considered according to the case. For different viruses with not the same obfuscation techniques and morphing level, threshold values may be selected completely different.

In the case of our study, distance values between each pairs of three versions of *Evol* virus are equal to zero. Anyway, if we use some other versions of this virus, maybe we have to choose a higher value to classify the other morphed instances of this virus. It is happened in the case of *Evul* virus variants. We should choose a threshold equal to **0.057** to classify the five versions of this virus. Highlighted cells show the distance less than threshold. Programs with highlighted distances are considered as morphed versions of each other. The other three instances of *Evul* are more different. Distance values for all versions of *Evul* are given in Table 12, separately. After we obtained the result, we analyze *Evul* Virus deeply, to find why other three instances have higher distance quantities. We find that in variants *Evul.d*, *Evul.g* and *Evul.f*, virus inserts *nop* instruction as dead code. For this reason, histograms of these versions are disparate and distance values are higher.

Table 10: List of experiment files

| Virus/Program | Description/Application |
|---|---|
| `Win32.Evol.a/b/c` | `Versions of virus Evol` |
| `Win32.Evul.`<br>`a/b/c/d/e/f/g/h` | `Versions of virus Evul` |
| `Lib.exe` | `Microsoft Visual Studio 9.0` |
| `Write.exe` | `Windows 7 WordPad` |
| `Wordconv.exe` | `Microsoft office 2007` |
| `Help.exe` | `Windows 7 command line help` |
| `Find.exe` | `Windows 7 command line find` |
| `Drvins32.exe` | `Kaspersky Antivirus 2010` |
| `Perlglob.exe` | `Mathworks MatLab R2008a` |
| `logoUI.exe` | `Windows 7` |

## 6 Conclusions and Future Works

This study shows that the frequency histogram of opcodes can be considered as a feature to classify the obfuscated versions of metamorphic viruses.

There are two major drawbacks with this method. Firstly, because a wide range of programs use some of the most common machine instructions and this method is highly depend on instruction frequency, is very difficult to choose appropriate threshold to decrease the risk of false positive. Secondly, it works only for a limited range of obfuscation techniques. Some metamorphic methods, such as instruction substitution and junk code insertion, can defeat this classification methodology.

For the future extension of this methodology, we can suggest some recommendations. One beneficial improvement is combining a weighted calculation to Minkowski-form distance metric. Some instructions, such as *mov*, *push*, *call* , and so on, are more using in programs. These kind of opcodes can be weighted to create a more precise distance metric.

Another valuable development is to modify methodology to overcome the other obfuscation techniques that is not included in current solution. Before we start the comparison step to find dissimilarity between two programs, we can analyze the programs and their subroutines to prune dead code inserted or solve the issue of mutation via exchangeable instructions to obtain a uniform minimal core. It is obvious that this pre-process may increase the complexity of the algorithm, especially in aspect of the time.

In addition, normalization of the histogram will eliminate the length of frequency vector. In this study, we had to normalize the histograms to achieve a threshold-based comparison for classification. However, a worthy study is to solve the threshold problem, eliminate the normalization of histogram, and compare the histograms in keeping with the number of opcodes, not according to proportion of frequency of opcodes.

Table 11: Distance values and classification based on threshold equal to **0.057**

| | Virus-Win32-Evol-a | Virus-Win32-Evol-b | Virus-Win32-Evol-c | Virus-Win32-Evul-8192-a | Virus-Win32-Evul-8192-b | Virus-Win32-Evul-8192-c | Virus-Win32-Evul-8192-e | Virus-Win32-Evul-8192-h | lib | write | wordconv | help | find | drvins32 | perlglob | LogonUI-til |
|---|---|---|---|---|---|---|---|---|---|---|---|---|---|---|---|---|
| Virus-Win32-Evol-a | 0.000 | 0.000 | 0.000 | 0.407 | 0.375 | 0.377 | 0.407 | 0.407 | 0.299 | 0.310 | 0.229 | 0.299 | 0.259 | 0.358 | 0.512 | 0.259 |
| Virus-Win32-Evol-b | 0.000 | 0.000 | 0.000 | 0.407 | 0.374 | 0.377 | 0.407 | 0.407 | 0.298 | 0.309 | 0.228 | 0.298 | 0.258 | 0.358 | 0.510 | 0.257 |
| Virus-Win32-Evol-c | 0.000 | 0.000 | 0.000 | 0.407 | 0.374 | 0.376 | 0.407 | 0.407 | 0.298 | 0.309 | 0.228 | 0.298 | 0.258 | 0.358 | 0.510 | 0.257 |
| Virus-Win32-Evul-8192-a | 0.407 | 0.407 | 0.407 | 0.000 | 0.036 | 0.032 | 0.000 | 0.004 | 0.450 | 0.477 | 0.316 | 0.402 | 0.358 | 0.216 | 0.460 | 0.384 |
| Virus-Win32-Evul-8192-b | 0.375 | 0.374 | 0.374 | 0.036 | 0.000 | 0.003 | 0.036 | 0.028 | 0.415 | 0.460 | 0.305 | 0.369 | 0.323 | 0.217 | 0.421 | 0.353 |
| Virus-Win32-Evul-8192-c | 0.377 | 0.377 | 0.376 | 0.032 | 0.003 | 0.000 | 0.032 | 0.024 | 0.429 | 0.465 | 0.310 | 0.381 | 0.335 | 0.222 | 0.434 | 0.353 |
| Virus-Win32-Evul-8192-e | 0.407 | 0.407 | 0.407 | 0.000 | 0.036 | 0.032 | 0.000 | 0.004 | 0.450 | 0.477 | 0.316 | 0.402 | 0.358 | 0.216 | 0.460 | 0.384 |
| Virus-Win32-Evul-8192-h | 0.407 | 0.407 | 0.407 | 0.004 | 0.028 | 0.024 | 0.004 | 0.000 | 0.456 | 0.490 | 0.320 | 0.407 | 0.358 | 0.227 | 0.456 | 0.387 |
| lib | 0.299 | 0.298 | 0.298 | 0.450 | 0.415 | 0.429 | 0.450 | 0.456 | 0.000 | 0.128 | 0.165 | 0.125 | 0.234 | 0.353 | 0.458 | 0.194 |
| write | 0.310 | 0.309 | 0.309 | 0.477 | 0.460 | 0.465 | 0.477 | 0.490 | 0.128 | 0.000 | 0.158 | 0.057 | 0.190 | 0.363 | 0.482 | 0.110 |
| wordconv | 0.229 | 0.228 | 0.228 | 0.316 | 0.305 | 0.310 | 0.316 | 0.320 | 0.165 | 0.158 | 0.000 | 0.147 | 0.211 | 0.242 | 0.368 | 0.170 |
| help | 0.299 | 0.298 | 0.298 | 0.402 | 0.369 | 0.381 | 0.402 | 0.407 | 0.125 | 0.057 | 0.147 | 0.000 | 0.195 | 0.321 | 0.437 | 0.135 |
| find | 0.259 | 0.258 | 0.258 | 0.358 | 0.323 | 0.335 | 0.358 | 0.358 | 0.234 | 0.190 | 0.211 | 0.195 | 0.000 | 0.313 | 0.505 | 0.146 |
| drvins32 | 0.358 | 0.358 | 0.358 | 0.216 | 0.217 | 0.222 | 0.216 | 0.227 | 0.353 | 0.363 | 0.242 | 0.321 | 0.313 | 0.000 | 0.473 | 0.302 |
| perlglob | 0.512 | 0.510 | 0.510 | 0.460 | 0.421 | 0.434 | 0.460 | 0.456 | 0.458 | 0.482 | 0.368 | 0.437 | 0.505 | 0.473 | 0.000 | 0.510 |
| LogonUI-til | 0.259 | 0.257 | 0.257 | 0.384 | 0.353 | 0.365 | 0.384 | 0.387 | 0.194 | 0.110 | 0.170 | 0.135 | 0.146 | 0.302 | 0.510 | 0.000 |

Table 12: Distance values for all 8 versions of *Evul*.

| | Virus-Win32-Evul-8192-a | Virus-Win32-Evul-8192-b | Virus-Win32-Evul-8192-c | Virus-Win32-Evul-8192-d | Virus-Win32-Evul-8192-e | Virus-Win32-Evul-8192-f | Virus-Win32-Evul-8192-g | Virus-Win32-Evul-8192-h |
|---|---|---|---|---|---|---|---|---|
| Virus-Win32-Evul-8192-a | 0.000 | 0.036 | 0.032 | 0.149 | 0.000 | 0.149 | 0.215 | 0.004 |
| Virus-Win32-Evul-8192-b | 0.036 | 0.000 | 0.003 | 0.161 | 0.036 | 0.161 | 0.179 | 0.028 |
| Virus-Win32-Evul-8192-c | 0.032 | 0.003 | 0.000 | 0.158 | 0.032 | 0.158 | 0.185 | 0.024 |
| Virus-Win32-Evul-8192-d | 0.149 | 0.161 | 0.158 | 0.000 | 0.149 | 0.000 | 0.056 | 0.153 |
| Virus-Win32-Evul-8192-e | 0.000 | 0.036 | 0.032 | 0.149 | 0.000 | 0.149 | 0.215 | 0.004 |
| Virus-Win32-Evul-8192-f | 0.149 | 0.161 | 0.158 | 0.000 | 0.149 | 0.000 | 0.056 | 0.153 |
| Virus-Win32-Evul-8192-g | 0.215 | 0.179 | 0.185 | 0.056 | 0.215 | 0.056 | 0.000 | 0.207 |
| Virus-Win32-Evul-8192-h | 0.004 | 0.028 | 0.024 | 0.153 | 0.004 | 0.153 | 0.207 | 0.000 |